\documentstyle[12pt]{article}

\def\gt{\ifmmode{{\mathchar"313E}}\else{$\rangle$}\fi}
\def\lt{\ifmmode{{\mathchar"313C}}\else{$\langle$}\fi}

\begin{document}
\title {Bi-continuum modelling of layered structures and
crystalline interfaces}
\author {Ma{\l}gorzata Sztyren\thanks{Department of Mathematics and Information Science,
Warsaw University of Technology, Pl. Politechniki 1, PL-00-661 Warsaw.
E--mail: emes@mech.pw.edu.pl}
}
\date{}
\maketitle
\hfill To the memory of Ekkehart Kr\"{o}ner\\

\input{epsf.tex}
\edef\undtranscode{\the\catcode`\_} \catcode`\_11
\newbox\box_tmp 
\newdimen\dim_tmp 
\def\jump_setbox{\aftergroup\after_setbox}
%
%
\def\resize
    #1
    #2
    #3
    #4
    {%
    \dim_r#2\relax \dim_x#3\relax \dim_t#4\relax
    \dim_tmp=\dim_r \divide\dim_tmp\dim_t
    \dim_y=\dim_x \multiply\dim_y\dim_tmp
    \multiply\dim_tmp\dim_t \advance\dim_r-\dim_tmp
    \dim_tmp=\dim_x
    \loop \advance\dim_r\dim_r \divide\dim_tmp 2
    \ifnum\dim_tmp>0
      \ifnum\dim_r<\dim_t\else
        \advance\dim_r-\dim_t \advance\dim_y\dim_tmp \fi
    \repeat
    #1\dim_y\relax
    }
\newdimen\dim_x    
\newdimen\dim_y    
\newdimen\dim_t    
\newdimen\dim_r    
\def\perc_scale#1#2{
  \def\after_setbox{%
    \hbox\bgroup
    \dim_tmp\wd\box_tmp \divide\dim_tmp100 \wd\box_tmp#1\dim_tmp
    \dim_tmp\ht\box_tmp \divide\dim_tmp100 \ht\box_tmp#2\dim_tmp
    \dim_tmp\dp\box_tmp \divide\dim_tmp100 \dp\box_tmp#2\dim_tmp
    \box\box_tmp 
\egroup}%
    \afterassignment\jump_setbox\setbox\box_tmp =
}%
{\catcode`\p12 \catcode`\t12 \gdef\PT_{pt}}
\def\hull_num{\expandafter\hull_num_}
\expandafter\def\expandafter\hull_num_\expandafter#\expandafter1\PT_{#1}
\def\find_scale#1#2{
  \def\after_setbox{%
    \resize\dim_tmp{100pt}{#1}{#2\box_tmp}%
    \xdef\lastscale{\hull_num\the\dim_tmp}\extra_complete}%
  \afterassignment\jump_setbox\setbox\box_tmp =
}
\def\scaleto#1#2#3#4{
  \def\extra_complete{\perc_scale{#3}{#4}\hbox{\box\box_tmp}}%
  \find_scale{#1}#2}
\let\xyscale\perc_scale
\def\zscale#1{\xyscale{#1}{#1}}
\def\yxscale#1#2{\xyscale{#2}{#1}}
\def\xscale#1{\xyscale{#1}{100}}
\def\yscale#1{\xyscale{100}{#1}}
\def\xyscaleto#1{\scaleto{#1}\wd\lastscale\lastscale}
\def\yxscaleto#1{\scaleto{#1}\ht\lastscale\lastscale}
\def\xscaleto#1{\scaleto{#1}\wd\lastscale{100}}
\def\yscaleto#1{\scaleto{#1}\ht{100}\lastscale}
\def\slant#1{
  \hbox\bgroup
  \def\after_setbox{%
    \box\box_tmp 
\egroup}%
  \afterassignment\jump_setbox\setbox\box_tmp =
}%
\def\rotate#1{
  \hbox\bgroup
  \def\after_setbox{%
    \setbox\box_tmp\hbox{\box\box_tmp}
    \wd\box_tmp 0pt \ht\box_tmp 0pt \dp\box_tmp 0pt
    \box\box_tmp
    \egroup}%
  \afterassignment\jump_setbox\setbox\box_tmp =
}%
\newdimen\box_tmp_dim_a
\newdimen\box_tmp_dim_b
\newdimen\box_tmp_dim_c
\def\plus_{+}
\def\minus_{-}
\def\revolvedir#1{
  \hbox\bgroup
   \def\param_{#1}%
   \ifx\param_\plus_ \else \ifx\param_\minus_
   \else
     \errhelp{I would rather suggest to stop immediately.}%
     \errmessage{Argument to \noexpand\revolvedir should be either + or -}%
   \fi\fi
  \def\after_setbox{%
    \box_tmp_dim_a\wd\box_tmp
    \setbox\box_tmp\hbox{%
     \ifx\param_\plus_\kern-\box_tmp_dim_a\fi
     \box\box_tmp
     \ifx\param_\plus_\kern\box_tmp_dim_a\fi}%
    \box_tmp_dim_a\ht\box_tmp \advance\box_tmp_dim_a\dp\box_tmp
    \box_tmp_dim_b\ht\box_tmp \box_tmp_dim_c\dp\box_tmp
    \dp\box_tmp0pt \ht\box_tmp\wd\box_tmp \wd\box_tmp\box_tmp_dim_a
    \kern \ifx\param_\plus_ \box_tmp_dim_c \else \box_tmp_dim_b \fi
    \box\box_tmp
    \kern -\ifx\param_\plus_ \box_tmp_dim_c \else \box_tmp_dim_b \fi
    \egroup}%
  \afterassignment\jump_setbox\setbox\box_tmp =
}%
\def\revolve{\revolvedir-}
\def\xflip{
  \hbox\bgroup
  \def\after_setbox{%
    \box_tmp_dim_a.5\wd\box_tmp
   \setbox\box_tmp
     \hbox{\kern-\box_tmp_dim_a \box\box_tmp \kern\box_tmp_dim_a}%
   \kern\box_tmp_dim_a
    \box\box_tmp
    \kern-\box_tmp_dim_a
    \egroup}%
  \afterassignment\jump_setbox\setbox\box_tmp =
}%
\def\yflip{
  \hbox\bgroup
  \def\after_setbox{%
    \box_tmp_dim_a\ht\box_tmp \box_tmp_dim_b\dp\box_tmp
    \box_tmp_dim_c\box_tmp_dim_a \advance\box_tmp_dim_c\box_tmp_dim_b
    \box_tmp_dim_c.5\box_tmp_dim_c
   \setbox\box_tmp\hbox{\vbox{%
     \kern\box_tmp_dim_c\box\box_tmp\kern-\box_tmp_dim_c}}%
   \advance\box_tmp_dim_c-\box_tmp_dim_b
   \setbox\box_tmp\hbox{%
     \lower\box_tmp_dim_c\box\box_tmp
}%
    \ht\box_tmp\box_tmp_dim_a \dp\box_tmp\box_tmp_dim_b
    \box\box_tmp
    \egroup}%
  \afterassignment\jump_setbox\setbox\box_tmp =
}%
\catcode`\_\undtranscode

\special{ps:}

\newcommand {\eqn}[1]{\begin{equation}#1\end{equation}}
\newcommand {\eqna}[1]{\begin{eqnarray}#1\end{eqnarray}}
\newcommand {\la}{\longrightarrow}

{\footnotesize {\bf Summary: }
The bi-continuum model composed of two interpenetrating and
dynamically coupled material continua is analysed as a simplified but
relatively accurate way to describe some physical phenomena
in crystalline solids. The essential novelty of our approach
consists in treating a crystalline medium as a bi-continuum,
even if the crystalline lattice is structurally
single-component. Particular attention is paid to the oscillatory
behaviour of solutions on the atomic level.
Starting from a discrete atomic chain, the basic formulation of
the bi-continuum model is derived. The essential features of the
model, including accuracy of the results
as functions of physical parameters, are discussed. 
}

\tableofcontents

\newpage
\section{Introduction}
\indent
    Discrete structural models such as atomic lattices or
stratified structures are capable of delivering, as a rule, highly
relialable and accurate results concerning a wide range of physical
phenomena in solid matter. Most frequently, however, to produce
such results, an atomic model has to be treated with the aid of
computer simulation techniques, what imposes severe restrictions
on the size of computed samples and leads to the lack of
generality in the results. Moreover, atomic models are sensitive
to details of the interatomic potentials; the
final reliability of results depends critically on the
reliability of the potentials used. In consequence,
simplified models which may produce analytical -- although
less accurate -- results are still highly desirable. The loss of
accuracy should not, however, be catastrophic.

    In the present paper we focus our attention on atomic
configurations in complex layered structures or in the
vicinity of planar or quasi-planar crystalline interfaces
\cite{MS Udine}.
In such situations, the configurations frequently exhibit
a sign-alternating behaviour 
\cite{Bristowe+Crocker,Rogula+Czerwinska 2001}
on atomic scale,
not reproducible by a simple classic continuum. We have found,
however, that such a behaviour
can be described by an appropriately chosen two-component continuum
model, called the bi-continuum.

   The concept of multi-continuum, understood as a material
medium on which several
displacement fields are defined, is intuitively evident when one
considers multicomponent media (like e.g. mixture of substances
with different physical properties
or the medium composed of electrons considered as one component and
heavy ions as the other one).

   The idea given in \cite{copenhagen}
was to introduce the multicontinuum
description for structurally complex media, even if such a medium
has only one component. The different displacement fields are
defined on different substructures. We shall show how the
idea of multicontinuum works in the case when multi=2.

   From the point of view of independent variables, the problem is
essentially one-dimensional. The corresponding variable will be
denoted by \(x\) with, according to the needs, appropriate indices
or diacritics. For the sake of simplicity, we perform our theoretical
constructions on linear atomic chains, although in the real
three-dimensional applications the "atoms" are intended to represent
whole crystalline planes or appropriate structural layers.

   In Sections 2 and 3 we state the problem for a homogeneous
atomic chain and examine the sign-alternating solutions near the
"interface". In Section 4 we analyse a composite linear chain,
consisting of two components. This provides the basis for construction
of the bi-continuum model in Section 5. Subsequently we solve
the interface problem in atomic and bi-continuum models, respectively,
compare the solutions, and discuss the resulting degree of
approximation.

\section{Homogeneous atomic chain}\label{Kajtek}
\indent
    Let us consider an atomic chain with interactions between the first and
second nearest neighbours. Let $u_{i}$  denote the displacement of the atom
number $i$ from its equilibrium position. By  $\kappa_{f}, n_{f}$
and $\kappa_{s}, n_{s}$ we denote the elastic constant and initial stretch
connected with shorter and longer bonds, respectively.\\
\vskip 0pt
\noindent \hfil\hbox{\epsffile{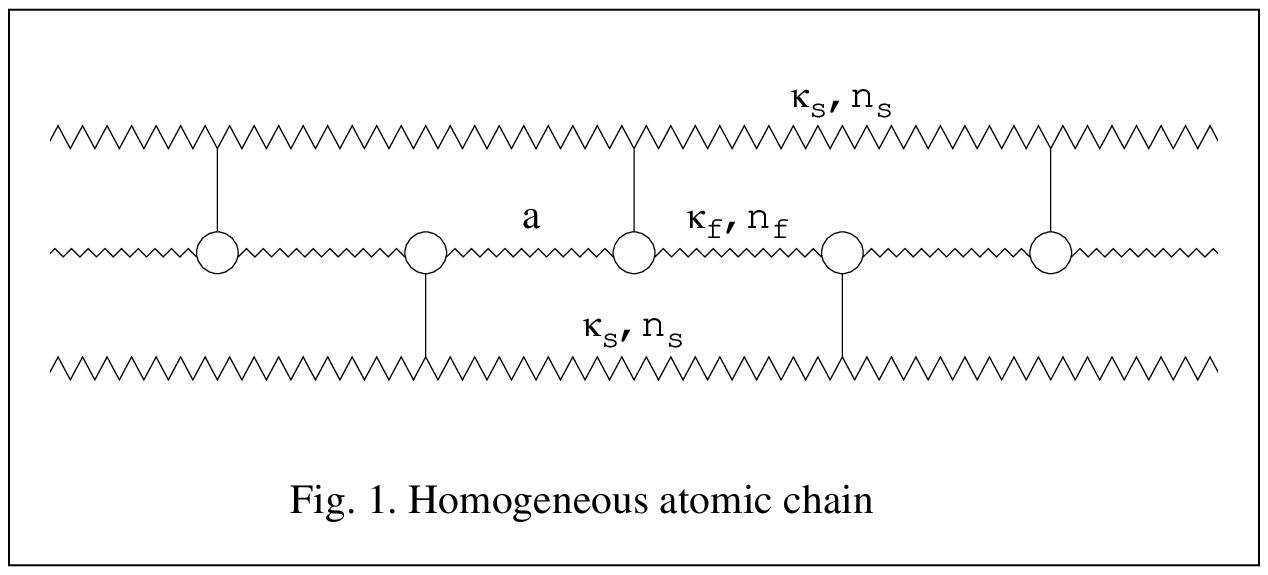}}\hfil\\
\vskip -0.5\baselineskip
The energy of a short bond is given by the formula
\eqn{W_{f}=\frac{1}{2}\kappa_{f}(\nabla_{1}u)^{2}+n_{f}\nabla_{1}u}
while that of a long one by
\eqn{W_{s}=\frac{1}{2}\kappa_{s}(\nabla_{2}u)^{2}+n_{s}\nabla_{2}u,}
where $\nabla_{1}$ and $\nabla_{2}$  are symbols of the central
difference operators acting on displacements of the first and second
neighbours, respectively.
The elastic energy of the chain equals the sum over all the bonds.
\par
      The state of equilibrium of a homogeneous chain is described by
the set of equations
\eqn{\kappa_{f}(u_{i-1}-2u_{i}+u_{i+1})+\kappa_{s}(u_{i-2}-2u_{i}+u_{i+2})=0,
           \label{ll3}}
where the index $i=1,2,3,\ldots$ .
The general solution to equations (\ref{ll3}) is given by the function
\eqn{ u_{i}=Az^{i}+Bz^{-i}+Ei+C}
with arbitrary constants $A, B, E, C$,  and
\eqn{ z = -\beta+\sqrt{\beta^{2}-1},\label{l3}}
where
\eqn{ \beta=\frac{\alpha}{2}+1,\ \  \alpha=\frac{\kappa_{f}}{\kappa_{s}}.}
\section{Atomic chain with "interface"\label{sec4}}

   Now, instead of a homogeneous chain we shall consider two
semi--infinite homogeneous parts (of possibly different materials)
connected by an `interface'
modelled by intermediate bonds of the third kind. The quantities related
to the intermediate bonds will be labelled by the superscript $^{o}$, while
the superscripts $^{-}$ and $^{+}$ will refer to the left and right
semi--infinite chains, respectively.\\
\par
For simplicity we assume that the parameter $\alpha$
has the same value for all three materials,
\eqn{ \frac{\kappa^{-}_{f}}{\kappa^{-}_{s}}=
      \frac{\kappa^{o}_{f}}{\kappa^{o}_{s}}=
      \frac{\kappa^{+}_{f}}{\kappa^{+}_{s}}=\alpha .}
Moreover, we assume that\\

a) the homogeneous chains are in equilibrium,\\

b) there are no forces at infinity.\\

The above conditions imply the following relation between
the initial stretch forces:
\eqn{ n_{f}+2n_{s}=0,}
which means that the sum of forces transmitted by any section vanishes.
\par
      The solution of such an interface problem is given by
\eqn{u_{j}=\left\{ \begin{array}{ll} \displaystyle
          \frac{\varepsilon_{-}}{1-z}z^{-j-1} & {\rm for } \ j=-1,-2,\dots\\
          \ & \ \\
          \displaystyle
          A-\frac{\varepsilon_{+}}{1-z}z^{j-1} & {\rm for } \ j=1,2,\dots\ ,
            \end{array} \right .\label{l9}}
where $z$  is defined by the formula (\ref{l3}). Displacements are calculated
relative to the asymptotic value at  $-\infty$. The asymptotic value
at $+\infty$  is given by
\eqn{A=\varepsilon_{o}+\frac{\varepsilon_{-}+\varepsilon_{+}}
                            {1-z} .}
The symbols  $\varepsilon_{-},\varepsilon_{o}$ and $\varepsilon_{+}$  denote
deformations and are expressed by material parameters as follows:
\eqn{\varepsilon_{-}=-T_{1}(\varepsilon_{o}+\frac{1}{\kappa_{s}^{o}}
                           (n_{s}^{o}-n_{s}^{-}))\ \ ,}
\eqn{\varepsilon_{o}=\frac{1}{\kappa_{s}^{o}}(2\beta-T_{1}-T_{2})^{-1}
     (T_{1}(n_{s}^{o}-n_{s}^{-})-T_{2}(n_{s}^{+}-n_{s}^{o}))\ \ ,
     \label{eps0}}
\eqn{\varepsilon_{+}=T_{2}(\varepsilon_{o}-\frac{1}{\kappa_{s}^{o}}
                           (n_{s}^{+}-n_{s}^{o}))\ \ ,}
where we have introduced the auxiliary notations
\eqn{T_{1}=(1-\frac{\kappa_{s}^{-}}{\kappa_{s}^{o}}(1+\frac{1}{z}))^{-1}\ \ ,}
\eqn{T_{2}=(1-\frac{\kappa_{s}^{+}}{\kappa_{s}^{o}}(1+\frac{1}{z}))^{-1}\ \ .}
The resulting solution is represented by sharp-bend vertices of the line in
Fig. 3.

   Let us notice that the number  $z$  is always negative.
It implies a strongly oscillating character of the solution given
by formula (\ref{l9}): the nearest neighbours are displaced in opposite
directions.
Thus every second atom is displaced in the same direction.
Hence, as a result
of the presence of the second neighbour interactions, we have obtained
a polarization of the chain, expressed as a relative displacement
of both substructures.

\section{Composite atomic chain}

To construct the bi-continuum model we start from introduction of two fields
of displacements: $v$ and $w$, each of them defined on one of the sub-chains.
\vskip 7pt
\noindent \hfil\hbox{\epsffile{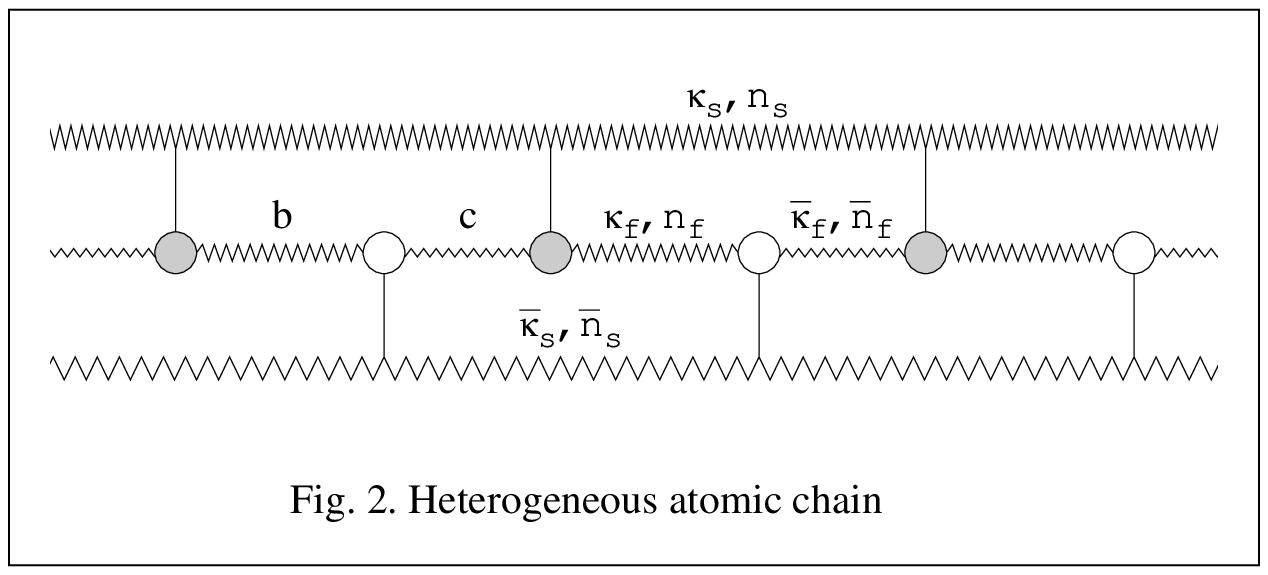}}\hfil\\
\vskip -0.5\baselineskip
We shall describe by the indices \(x\) and \(\bar{x}\) the position of "black"
and "white" atoms supporting the fields \(v\) and \(w\), respectively. The
description may be applied to two-component structures as well as to
one-component but dimerized ones. Altogether, in the translationally
invariant case we
have four classes of bonds. Two bonds
between the first neighbours are charcterised by parameters \(\kappa_f,\ n_f\)
and \(\bar{\kappa}_f,\ \bar{n}_f\), while two other bonds, between the second
neighbours,
are characterised by \(\kappa_s,\ n_s\) and \(\bar{\kappa}_s,\ \bar{n}_s\).
If there are no forces at infinity,
then the sum of forces transmitted by any section vanishes,
\eqna{n_s+n_f+\bar{n}_s=0,\nonumber\\
      n_s+\bar{n}_f+\bar{n}_s=0.}
This fact implies, in turn, that initial stretches of bonds between
the nearest neighbours equal each other:
\eqn{n_f=\bar{n}_f.\label{nf}}
The symbol \(a\) denotes now the distance between the closest atoms of
the same colour (which corresponds to \(2a\) in the previous sections),
\(b\) and \(c\) denote the distances between atoms of different colours.
These quantities are connected by the relation
\eqn{b+c=a.}
The energy of the chain can be composed of 4 parts, each one connected with
one kind of bonds,
\eqn{W=W_s+W_{\bar{s}}+W_f+W_{\bar{f}},}
where
\eqna{W_s=\sum(\frac{1}{2}\kappa_s(v_x-v_{x-a})^2+
   n_s(v_x-v_{x-a})),\nonumber \\
   W_{\bar{s}}=\sum(\frac{1}{2}\bar{\kappa}_s(w_{\bar{x}+a}-w_{\bar{x}})^2+
   \bar{n}_s(w_{\bar{x}+a}-w_{\bar{x}}))),\nonumber \\
   W_f=\sum(\frac{1}{2}\kappa_f(v_x-w_{\bar{x}})^2+
   n_f(v_x-w_{\bar{x}})),\nonumber \\
   W_{\bar{f}}=\sum\frac{1}{2}\bar{\kappa}_f(w_{\bar{x}}-v_{x-a})^2+
   \bar{n}_f(w_{\bar{x}}-v_{x-a})).
   \label{wwww}}
The last two parts of the energy contain terms combining values of different
fields at different points. We shall express these terms with the help of
fields \(\bar{v}\) and \(\bar{w}\) which interpolate the fields
\(v\) and \(w\) to the whole chain. By definition,
\eqn{\bar{v}_{\bar{x}}=\frac{\kappa_fv_x+\bar{\kappa}_fv_{x-a}}
                {\kappa_f+\bar{\kappa}_f},\ \ \ \bar{v}_x=v_x}
and
\eqn{\bar{w}_x=\frac{\bar{\kappa}_fw_{\bar{x}+a}+\kappa_fw_{\bar{x}}}
            {\kappa_f+\bar{\kappa}_f},\ \ \ \
           \bar{w}_{\bar{x}}=w_{\bar{x}}.}
Using these interpolating fields, we can express the energy of the first
neighbour bonds by
\eqn{W_f=\sum(\frac{1}{2}\kappa_f[(v_x-\bar{w}_x)+
        \frac{\bar{\kappa}_f}{\kappa_f+\bar{\kappa}_f}
        (w_{\bar{x}+a}-w_{\bar{x}})]^2+n_f[(v_x-\bar{v}_x)+
        \frac{\bar{\kappa}_f}{\kappa_f+\bar{\kappa}_f}
        (w_{\bar{x}+a}-w_{\bar{x}})]),}
\eqn{W_{\bar{f}}=\sum(\frac{1}{2}\kappa_{\bar{f}}
      [(w_{\bar{x}}-\bar{v}_{\bar{x}})+
        \frac{\kappa_f}{\kappa_f+\bar{\kappa}_f}
        (v_x-v_{x-a})]^2+\bar{n}_f[(w_{\bar{x}}-\bar{v}_{\bar{x}})+
        \frac{\kappa_f}{\kappa_f+\bar{\kappa}_f}
        (v_x-v_{x-a})]).\label{ww}}
All the differences are now expressed either by two fields at the same
point, or by one field at different points.

\section{The bi-continuum}

We arrive at the corresponding bi-continuum expression for energy
by appropriate truncations of the Taylor series. As the simplest
possibility let us consider the correspondence
\eqna{(1)\ \ \ \nabla{f} \la af' \ ,\nonumber\\
      (2)\ \ \sum \la\ \ \frac{1}{a}\int\ ,\nonumber\\
      (3)\ \ \ \bar{f} \la f.\ \ \ \ \ \ \ \
      \label{cor0}}
The symbol $f$ represents here an arbitrary function, the dash denotes
the derivative with respect to \(x\), and $a$ -- the lattice parameter.
The set of rules
(\ref{cor0}) will be referred to as the correspondence of order 0.
We shall consider also
another set of rules, called correspondence of order 1, which takes
into account the microstructure of the elementary cell of our composite
chain. Namely, instead of rule (3) in (\ref{cor0}), we take
\eqna{(3a)\ \ \ \bar{v}\la\ \ v+\frac{\bar{\kappa}}
      {\kappa_f+\bar{\kappa}_f}v',\nonumber\\
      (3b)\ \ \bar{w}\la\ w-\frac{\bar{\kappa}}
      {\kappa_f+\bar{\kappa}_f}w',}
where 
\eqn{\bar{\kappa}=\kappa_fc-\bar{\kappa}_fb.\label{barka}}
If \(\bar{\kappa}_f=\kappa_f\) and
\(b=c\), then both the correspondence rules are equivalent.
Let us note that the transition to correspondence rules of order 1 does not
augment the order of equations. 
\par
    Using the correspondence rule of order 0 one obtains the following
formulae for energies of the corresponding bonds:
\eqn{W_s=\int( \frac{a}{2}\kappa_sv'^2+n_sv')dx,}\nonumber

\eqn{W_{\bar{s}}=\int(\frac{a}{2}\bar{\kappa}_sw'^2+\bar{n}_fw')dx,}\nonumber

\eqn{W_f=\frac{1}{2a}\int[\kappa_f
  (v-w+\frac{\bar{\kappa}_f}{\kappa_f+\bar{\kappa}_f}aw')^2+
  an_f(v-w+\frac{\bar{\kappa}_f}{\kappa_f+\bar{\kappa}_f}aw')]dx,}\nonumber

\eqn{W_{\bar{f}}=\frac{1}{2a}\int[\bar{\kappa}_f
  (w-v+\frac{\kappa_f}{\kappa_f+\bar{\kappa}_f}av')^2+
  an_f(w-v+\frac{\kappa_f}{\kappa_f+\bar{\kappa}_f}av')]dx.}
In the last expression the equality (\ref{nf}) has been used. The energy
connected with the n.n. bonds may be rewritten in the form
\eqna{W_f+W_{\bar{f}}=\frac{1}{2a}\int[(\kappa_f+\bar{\kappa}_f)
  (v-w)^2+a^2\frac{\kappa_f\bar{\kappa}_f}{(\kappa_f+
  \bar{\kappa}_f)^2}\nonumber
  (\bar{\kappa}_fw'^2+\kappa_fv'^2)\\
  +\frac{an_f}{\kappa_f+
  \bar{\kappa}_f}(\bar{\kappa}_fw'+\kappa_fv')]dx
  +boundary\ term.}
The correspondence rule of order 1 introduce definite corrections to
the above formulae. These corrections result in
\par
a) an additional coupling term of the form \((v-w)(v'+w')\),
\par
b) renormalisation of coefficients in other terms.

   In consequence, the energy density of the homogeneous bi-continuum
takes the form
\eqn{w_d=\frac{1}{2}ev'^2+\frac{1}{2}gw'^2+\bar{\kappa}(v-w)(v'+w')+
         \frac{1}{2}h(v-w)^2,}
with the phenomenological coefficients. One can connect them with parameters
of the discrete model by the following equalities:
\eqna{e=a\kappa_s+a\bar{\kappa}_f(\frac{\kappa_f-\bar{\kappa}}
       {\kappa_f+\bar{\kappa}_f})^2,\nonumber\\
      g=a\bar{\kappa}_s+a\kappa_f(\frac{\bar{\kappa}_f-\bar{\kappa}}
       {\kappa_f+\bar{\kappa}_f})^2,\nonumber\\
      h=\frac{{\kappa}_f+\bar{\kappa}_f}{a},\label{pheno}}
and \(\bar{\kappa}\) is given by the formula (\ref{barka}).

\par
      The equations of static equilibrium for a homogeneous medium
have the form
\eqna{ev''-2\bar{\kappa}w'-h(v-w)=0,\nonumber\\
      gw''+2\bar{\kappa}v'+h(v-w)=0,}
and, without forces at infinity, have the general solution
\eqna{v=\left\{ \begin{array}{ll}
        Ae^{-\zeta x}+C, & {\rm for} \ x{\gt}0,\\
        \ & \ \\
        Be^{\zeta{x}} & {\rm for } \ x{\lt}0, \nonumber\\
        \end{array} \right\} \nonumber\\
      w=\left\{ \begin{array}{ll}
         Ak_1e^{-\zeta x}+C, & {\rm for} \ x{\gt}0,\\
        \ & \ \\
        Bk_2e^{\zeta{x}} & {\rm for } \ x{\lt}0.
        \end{array} \right\}
         \label{solu}}
with the exponent coefficient expressed by material constants according to
the formula
\eqn{\zeta^2=\frac{h(e+g)-4\bar{\kappa}^2}{eg}.}
The coefficients connecting amplitudes of fields \(v\) and \(w\) equal
\eqn{k_1=\frac{h-\zeta^2e}{h-2\zeta\bar{\kappa}},\ \ \
     k_2=\frac{h-\zeta^2e}{h+2\zeta\bar{\kappa}}  .}
In the case of exact inversion symmetry, when \(\kappa_s=\bar{\kappa}_s, b=c, 
\kappa_f=\bar{\kappa}_f\) and then \(\bar{\kappa}=0\),
\eqn{\zeta^2=\frac{2h}{e},\ \ k_1=k_2=-1.\label{dz}}
In the next section we shall consider such a symmetric case.

\section{Comparison between bi-continuum and discrete solutions}
   The solution of the bi-continuum interface problem may be constructed
from functions
(\ref{solu}), with appropriate values of constants \(A, B, C\). 
The simplest way to match this solution to its discrete counterpart is to
request that values of corresponding branches of functions \(u, v\) and
\(w\) coincide at the atoms nearest to the interface. Two such
conditions
plus coincidence of asymptotic values will be sufficient to calculate
\(A, B\) and \(C\). In this way, making use of (\ref{eps0}), we have
\eqn{C=\frac{T_1(n^0_s-n^-_s)}{\kappa^0_s(\beta-T_1)},}
and, for the other constants
\eqn{A=B=-\frac{\varepsilon_1}{1-z}e^{\zeta a/4}.}

\newpage
\noindent
The bi-continuum fields \(v\) and \(w\) are plotted in Fig. 3.
\vskip 7pt
\noindent \hfil\hbox{\epsffile{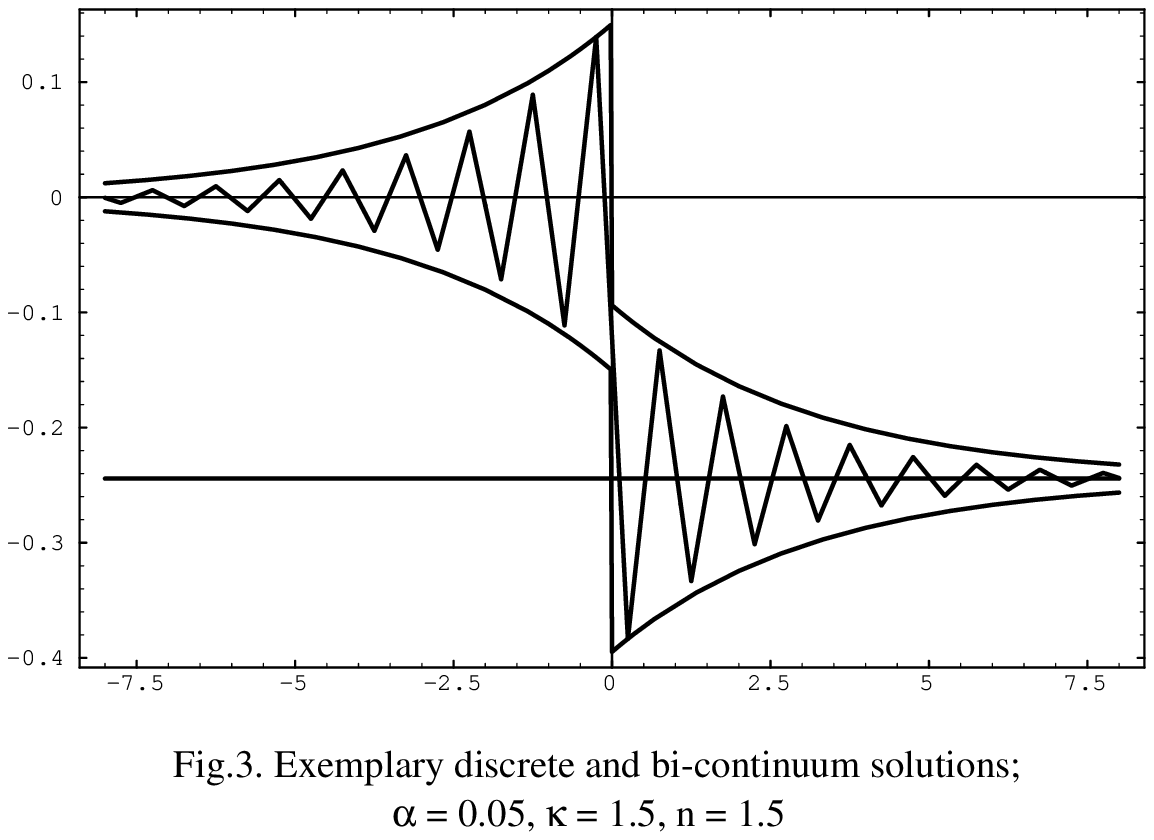}}\hfil\\
\vskip -0.5\baselineskip
\noindent
The upper branches correspond to the field \(w\).
The values of
discrete function \(u\) correspond to the sharp-bends
of the broken line.
\vskip 7pt
\noindent \hfil\hbox{\epsffile{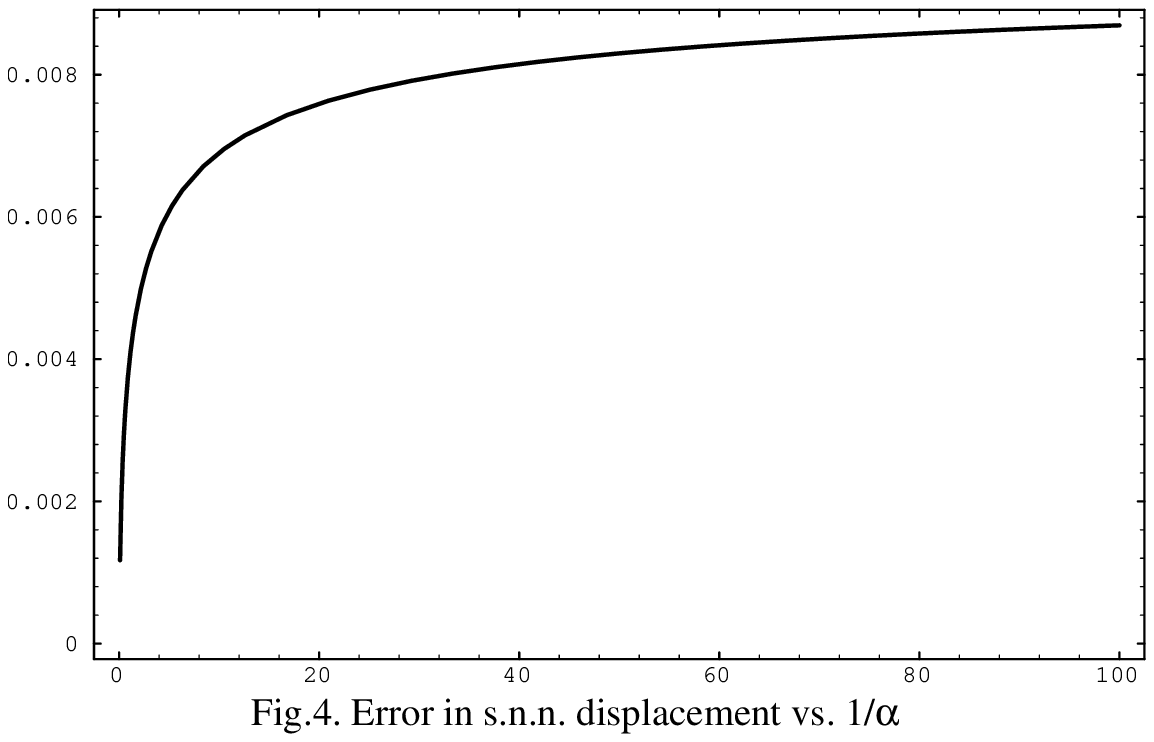}}\hfil\\
\vskip -0.5\baselineskip
   To examine the accuracy of the bi-continuum model,
the error in value of fields \(v\) and \(w\) at the distance \(3a/4\)
from the interface, equivalent to the second nearest neighbours,
is plotted vs. \(1/\alpha\) in Fig. 4.
The error corresponding to \(\alpha=0.05\) equals about 4 percent of
the lattice constant.
Note that, under the assumed correspondence rules, the error
does not change the sign.

   Another way to compare the atomic and bi-continuum models is to
analyse the relation between the local dimerisation ranges
or, equivalently, between the exponents in both models.
Let us note that, since the symbol \(a\) denotes the distance between
the second n.n., one should calculate the discrete exponent \(\lambda\)
from the equation
\eqn{z^2=e^{-\lambda a},}
which, according to (\ref{l3}), implies the dependence of \(\lambda\)
on \(\alpha\). On the other hand, for the bi-continuum model, by
making use of equations (\ref{dz}) and (\ref{pheno}), one obtains
\eqn{\zeta^2=\frac{16}{a^2}\frac{\alpha}{4+\alpha}.\label{ze}}
The above equation implies \(0{\lt}a\zeta{\lt}4\). 
\vskip 7pt
\noindent \hfil\hbox{\epsffile{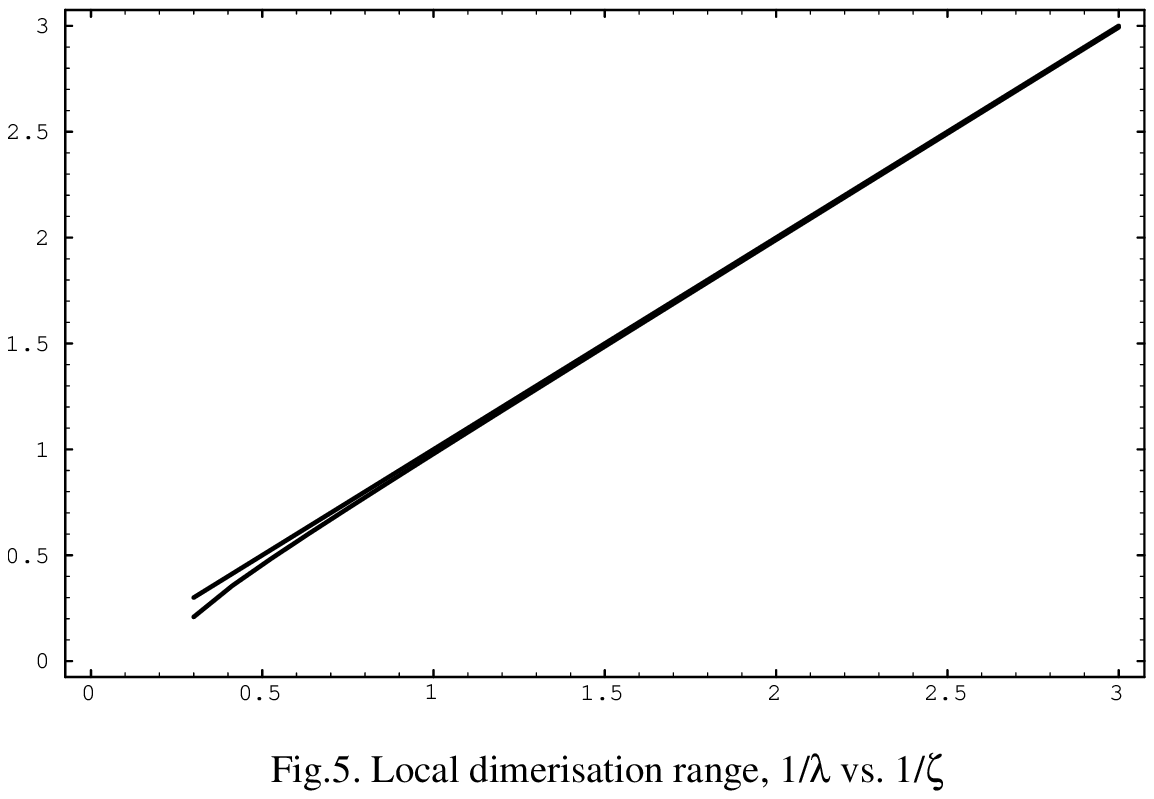}}\hfil\\
\vskip -0.5\baselineskip

\noindent
The dependence of \(1/\lambda\) on \(1/\zeta\), is plotted in Fig. 5
as the lower curve to be compared with the line \(\lambda=\zeta\). A
substantial difference is present only at distances smaller than
the lattice constant \(a\).
\newpage
\noindent
To enable a more precise estimation of the error, the difference
\(1/\lambda - 1/\zeta\)
is plotted vs. \(1/\alpha\) in Fig. 6.
\vskip 7pt
\noindent \hfil\hbox{\epsffile{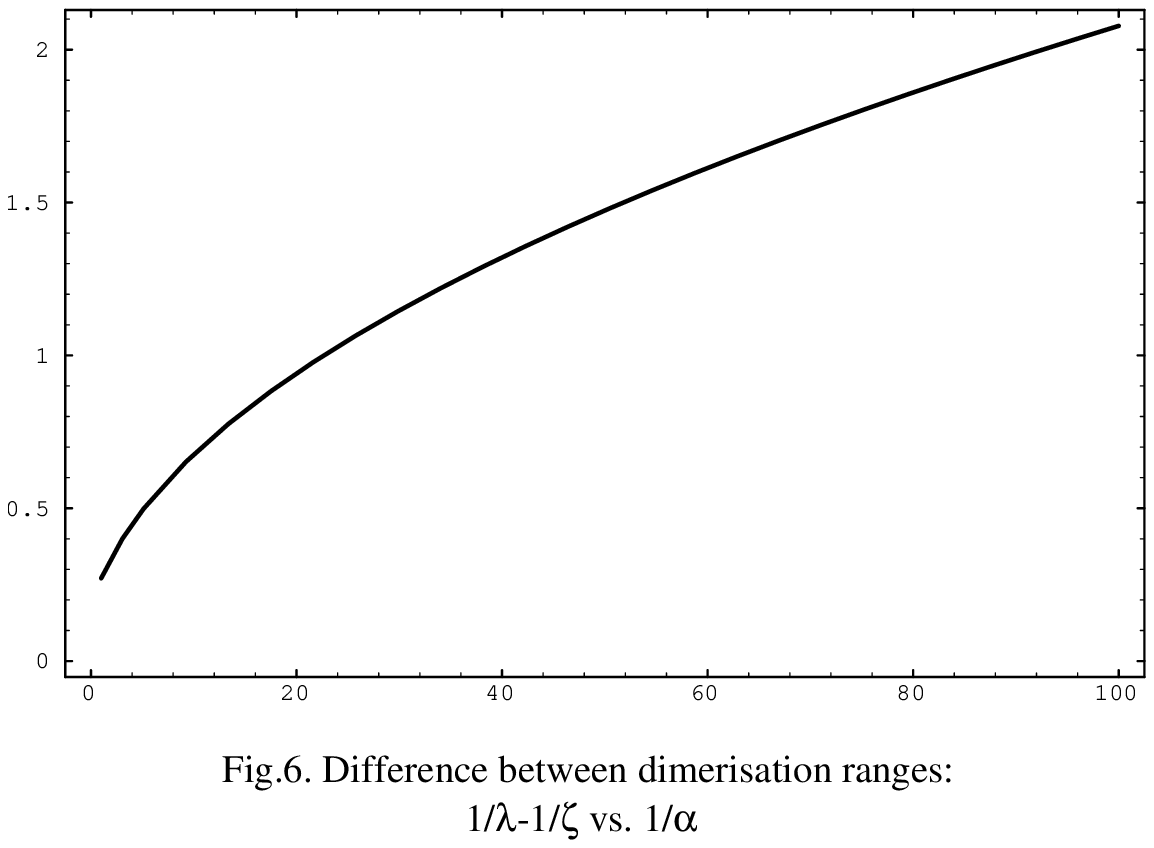}}\hfil\\
\vskip -0.5\baselineskip
\section{Final remarks}
   We have presented a method of construction of the bi-continuum model
based on a heuristic system - a linear atomic chain. One may, however,
forget about the derivation, and consider the obtained equations as
phenomenological ones. Generally, it is not necessary to relate the
phenomenological coefficients to the parameters of an atomic model.
The fairly good agreement between the bi-continuum and atomistic solutions
can be treated as an argument in behalf of our model, at least in the
sense that we have not made gross errors.

   The internal structure of the "atoms" in the chain may be complex --
they need not be interpreted as material points. As an important example
one can mention atomic planes of 3D crystalline structures, in particular
the oxygen-copper planes, or even the octahedral layers in high
temperature superconducing materials \cite{Rogula+Czerwinska 2001}.
   If our "atoms" are planes or layers \cite{Kroner 67},
one should replace, in the
expression for
energy, the scalar quantities by apropriate vectors and matrices.
And the construction will run analogically, under the condition that one
takes into account the interactions of the first and second n.n.
between planes or layers.
   Similarly one can consider layered structures in the form
of arrays of Josephson junctions. Such systems occur in a natural
way in superconducting crystals, \cite{Walker+Luettmer 96}
or else they can be shaped artificially
in a technological way.

   How important are the planar interfaces? There are many cases
where they are important by themselves. It is expedient to notice
that (what we call) the "planar" interfaces need not
be strictly planar. It is sufficient that the curvatures of the
corresponding surfaces should be not too high as compared with the
characteristic exponents (both of dimension \(cm^{-1}\),
inverse length).

   The interfaces can also affect the bulk properties of materials.
It concerns the materials which contain a lot of interface area per
unit volume. Also this property can be expressed by a parameter of
dimension of inverse length which is particularly high for
nanostructured materials.

\section{Acknowledgements}
This work was supported by the State Committee for Scientific Research
(Poland) under grant No. 7 TO7A 010 16.\\
\noindent
The paper has been presented at the International Symposium
on Structured Media, Pozna\a'{n}, 16-21 September 2001

\thebibliography{12}
\bibliography{}
\end{document}